\documentclass[12pt,a4paper]{article}
\usepackage{amsmath}
\usepackage{graphicx}% Include figure files
\usepackage{bm}% bold math

\begin{document}
\begin{center}
{\Large \textbf{Signature of \textit{f}-electron conductance in  $\alpha$-Ce single-atom contacts}}\\
\vspace{6mm}
Sebastian Kuntz$^{1}$, Oliver Berg$^{1}$, Christoph S\"urgers$^{1\ast}$, and Hilbert v. L\"ohneysen$^{1,2}$\\
\vspace{6mm}
$^1$Karlsruhe Institute of Technology, Physikalisches Institut, P.O. Box 6980, D-76049 Karlsruhe, Germany\\
$^2$Karlsruhe Institute of Technology, Institut f\"ur Festk\"orperphysik, \\P.O. Box 3640, D-76021 Karlsruhe, Germany\\
$^{\ast}$e-mail: christoph.suergers@kit.edu
\end{center}
\vspace{6mm}
\noindent
\textbf{Cerium is a fascinating element exhibiting, with its different phases, long-range magnetic order and superconductivity in bulk form. The coupling of the 4\textit{f} electron to \textit{sd} conduction electrons and to the lattice is responsible for unique structural and electronic properties like the isostructural first-order solid-solid transition from the cubic $\gamma$ phase to the cubic $\alpha$ phase, which is accompanied by a huge volume collapse of 14 \%. While the $\gamma - \alpha$ phase transition has been investigated for decades, experiments aiming at disentangling the 4\textit{f} contribution to the electric conductance of the different phases have not been performed. Here we report on the strongly enhanced conductance of single-atom Ce contacts. By controlling the content of $\alpha$-Ce employing different rates of cooling, we find a strong correlation between the fraction of $\alpha$-Ce and the magnitude of the last conductance plateau before the contact breaks. We attribute the enhanced conductance of $\alpha$-Ce to the additional contribution of the 4\textit{f} level.}

%Among the elements the rare-earth metal cerium takes a unique position due to one electron filling the \textit{f} orbital while the $5d^1$ and $6s^2$ electrons contribute to the conduction band of hybridized $spd$ bands. The \textit{f} electron plays a key role in the first-order structural phase transition between the face-centered cubic (fcc) $\gamma$ phase and the fcc $\alpha$ phase. At zero pressure, the $\gamma$ phase has a larger volume and shows a Curie-Weiss behaviour of the magnetic susceptibility due to localized but disordered 4\textit{f} moments [???]. Below 250 K, it first transforms into the double hexagonal close-packed (dhcp) $\beta$ phase and below 106-116 K into the Pauli paramagnetic $\alpha$ phase where the 4\textit{f} electrons are delocalized, participate in the bonding and give rise to a concomitant huge volume collapse of 14 \%. Upon heating, the $\alpha \rightarrow \gamma$ transformation starts at 174-178 K exhibiting a strong hysteresis between cooling and heating \cite{gschneidner_effects_1962}. Hence, the $\alpha \rightleftharpoons \gamma$ phase transition is a model system for a phase transition due to the delocalization of localized electrons under pressure or decrease of temperature.  
\clearpage
Cerium is perhaps the elemental material that exhibits the most pronounced configurational changes. Under ambient pressure, Ce is in the fcc phase ($\gamma$-Ce) in the configuration [Xe](6s5d)$^3$4f$^1$, with the 4$f$ electron strongly localized, and exhibits Curie-Weiss-type paramagnetism \cite{lawson_concerning_1949,koskenmaki_chapter_1978}. However, below $\sim$ 200\,K the ground state of $\alpha$-Ce is [Xe](6s5d)$^4$4f$^0$ and the 4$f$ electron is delocalized \cite{koskimaki_heat_1975}. $\alpha$-Ce has the same fcc structure as $\gamma$-Ce, but the lattice constant $a$ changes from 5.15 to 4.85\,{\AA}. Noteworthy, $\alpha'$-Ce, a high pressure variant of the $\alpha$ phase, is even superconducting with $T_c$ = 1.7 K \cite{wittig_superconductivity_1968,Loa_lattice_2012}. Cerium is thus a paradigm of the interplay of magnetism and superconductivity. 

The early proposal describing the $\gamma \rightarrow \alpha$ phase transition by the promotion of the \textit{f} electron to the \textit{sd} conduction band \cite{lawson_concerning_1949,schuch_structure_1950} was found to be in disagreement with subsequent experiments \cite{gustafson_positron_1969,kornstadt_investigation_1980,podloucky_band_1983,patthey_low-energy_1985} and, furthermore, not confirmed by band-structure calculations \cite{min_total-energy_1986}. Instead, a delocalization of the 4\textit{f} electron into a 4\textit{f} band in $\alpha$-Ce was suggested pointing towards an orbitally selective Mott transition (MT) \cite{gustafson_positron_1969,johansson1974}. 

The nature of the $\gamma-\alpha$ transition, which can be tuned at ambient temperature by hydrostatic pressure, is still under  debate \cite{held_cerium_2001,de_medici_mott_2005,lanata_2013}. The issue is complicated by the existence of an intervening ($320 \geq T \geq 170$ \,K) dhcp phase ($\beta$-Ce). The $\beta$ phase has similar electronic properties as the $\gamma$ phase, with localized 4\textit{f} moments that order antiferromagnetically below 12.5 K \cite{wilkinson_neutron_1961,gibbons_magnetic_1987}. This has been confirmed by density functional theory (DFT) in the local density approximation taking the onsite Hubbard interaction  into account (LDA + U) \cite{amadon_ensuremathgamma_2008}. The $\gamma - \alpha$ transition proceeds much faster than the $\gamma - \beta$ transition. Since the transitions between these phases are of first order, it is very difficult to obtain single-phase Ce modifications \cite{wilkinson_neutron_1961,koskimaki_preparation_1974}. 

\begin{figure}
\centerline{\includegraphics[width=\columnwidth,clip=]{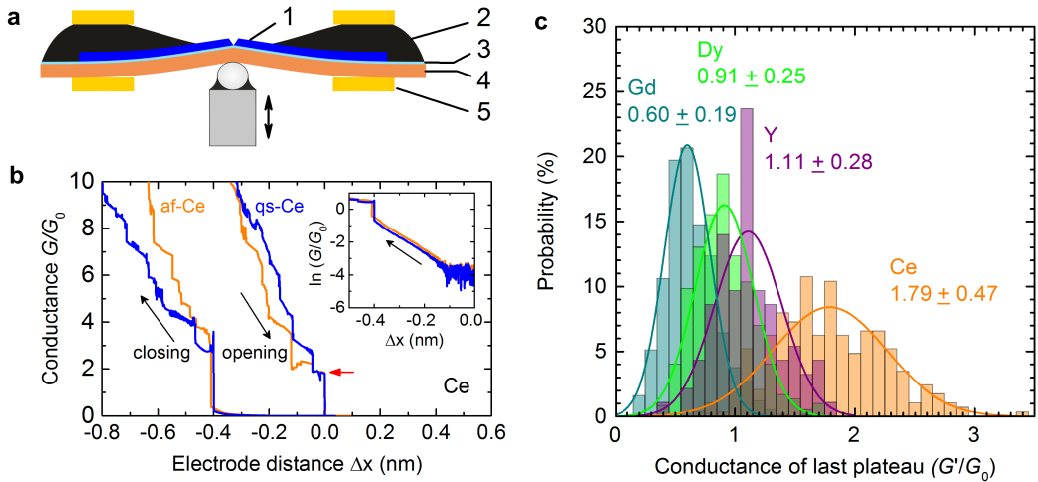}}
\caption[]{\textbf{Conductance of rare-earth nanocontacts. a,} Schematic of a mechanically-controlled break junction. 1: notched Ce wire, 2: Stycast epoxy, 3: insulating durimide layer, 4: bendable phosphor-bronze substrate, 5: countersuppports. \textbf{b,} Conductance $G$ vs. electrode distance ($\propto$ piezo voltage) of a cerium break-junction at 4.2 K. Red arrow marks the conductance of the last plateau before breaking. Inset shows semi-logarithmic plots of the conductance vs. electrode distance while closing the contact. The linear behaviour ln$(G/G_0) \propto \Delta x$ observed below the jump to contact at $\Delta x \approx -0.4$ \, nm is characteristic for electron tunneling. \textbf{c,} Histograms of the conductance $G'$ of the last plateau of Gd, Dy \cite{muller_switching_2011}, Y, and Ce  nanocontacts with a bin size of 0.1 $G'/\textnormal{G}_0$. For each histogram, the total probability integrated over all events is 100 \%. Solid lines represent a Gaussian distribution fit to the data with the mean value $\bar{G'}$ and standard deviation $\sigma$ indicated.}
\label{fig1}
\end{figure}

Although the structural and electronic properties of $\alpha$-, $\beta$-, and $\gamma$-Ce have been studied experimentally and theoretically for decades, the different contributions of \textit{s}, \textit{d}, and \textit{f} electrons to the total conductance have not been resolved. Mechanically controlled break junctions (MCBJ) offer the possibility to repeatedly open and close a contact established in a thin metallic wire (yet of macroscopic dimensions) and approach the quantum regime where the magnitude of the conductance is of the order of a few  conductance quanta G$_0 = 2 e^2/h$ \cite{agrait_quantum_2003}. Here the conductance $G$ exhibits plateau-like features when the contact is gradually opened mechanically, intercepted by sharp steps to lower $G$. It has been demonstrated for many different metals that the transport at the last plateau before breakage is due to current flow through a single atom \cite{agrait_quantum_2003}. Further increasing the distance between the electrodes yields vacuum tunneling through the opened contact as evidenced by the exponential distance dependence of $G \ll {\rm G}_0$. The conductance on the last plateau, on the order of $G_0$, depends on the number of atomic valence orbitals at or close to the Fermi level $E_{\rm F}$ and on the transmission coefficients of the orbitals \cite{scheer_signature_1998}. Therefore, this technique is well suited to investigate atomic-size contacts of elemental metals with different electron configurations like cerium. We note in passing that the situation in metals differs distinctly from that in semiconductors. In the latter, conductance quantization, i.e., $G = n$ $G_0$ with $n$ integer, has been observed because the inverse Fermi wave number $k_{\rm F}^{-1}$ is much larger than the interatomic distance \textit{a} due to the small conduction-electron density. In metals, on the other hand, $k_{\rm F}^{-1} \approx a$ leading to a strong intertwining of electronic properties and atomic structure in the contact. Here we report on the conductance of Ce single-atom contacts of MCBJ on the last plateau, i.e., single-atom contacts, obtained from polycrystalline wires (Fig. \ref{fig1}a). 

Figure \ref{fig1}b shows the conductance $G$ in units of $G_0$ vs. the electrode distance $\Delta x$ of cerium contacts at 4.2 K. Here, $\Delta x$ was determined from conductance measurements in the tunneling regime and the distance zero was arbitrarily set to the onset of conductance upon closing the contact signaled by a discontinuous jump from $G \ll {\rm G}_0$ to $G \approx {\rm G}_0$. The curves are typical of conductance curves of many elemental metals \cite{agrait_quantum_2003}. Upon stretching the contact, the conductance decreases by showing several steps due to the structural relaxation of the material and reformation of the atomic structure at the neck until finally a last plateau (Fig. \ref{fig1}b, red arrow) is reached before the contact eventually breaks and the conductance jumps to zero. Usually this behaviour is evaluated statistically on a large number of curves to reveal the most frequently occurring conductance values \cite{agrait_quantum_2003}. 

Here we focus on the conductance \textit{G'} of the last plateau before breaking at 4.2 K, characteristic of the conductance of a single-atom contact, for some rare-earth metal contacts in Fig. \ref{fig1}c. We find, again as usual, a broad distribution of conductances. While for ferromagnetic Gd, ferromagnetic dysprosium \cite{muller_switching_2011}, and nonmagnetic yttrium we observe a distribution of \textit{G'} with a maximum at $\bar{G'} = 0.6\, G_0$, $\bar{G'} = 0.91\, G_0$, or $\bar{G'} = 1.11 \, G_0$, respectively, a broader distribution with a maximum at $\bar{G'} = 1.79 \, G_0$ is observed for this particular contact of cerium. The values of $\bar{G'}$ for Dy and Y are close to $G_0$ as likewise observed for 3\textit{d} transition metals \cite{agrait_quantum_2003}. The conductance for Gd is in agreement with a recent investigation of lithographically prepared Gd MCBJ, where the conductance histogram - taking into account all plateaus observed below 20 $G_0$ - revealed a maximum at 0.75\, $G_0$ \cite{olivera_electronic_2016}. The low conductance of Gd is attributed to the hybridization between $s$ and $p_z$ conduction channels which reduces the conductance of the pure $s$ channels as inferred from DFT calculations. The electronic transport through atomic contacts is usually considered ballistic where the conductance is expressed by transport channels with a certain transmission of the electronic wave function in the Landauer-B\"uttiker theory \cite{agrait_quantum_2003}. The strongly enhanced conductance of Ce single-atom contacts compared to transition metals and to the rare-earth metals Gd and Dy with localized 4$f$ conduction electrons suggests that additional transport channels contribute to the total conductance of Ce which are attributed to the 4\textit{f} orbital.

\begin{figure}
\includegraphics[width=\columnwidth,clip=]{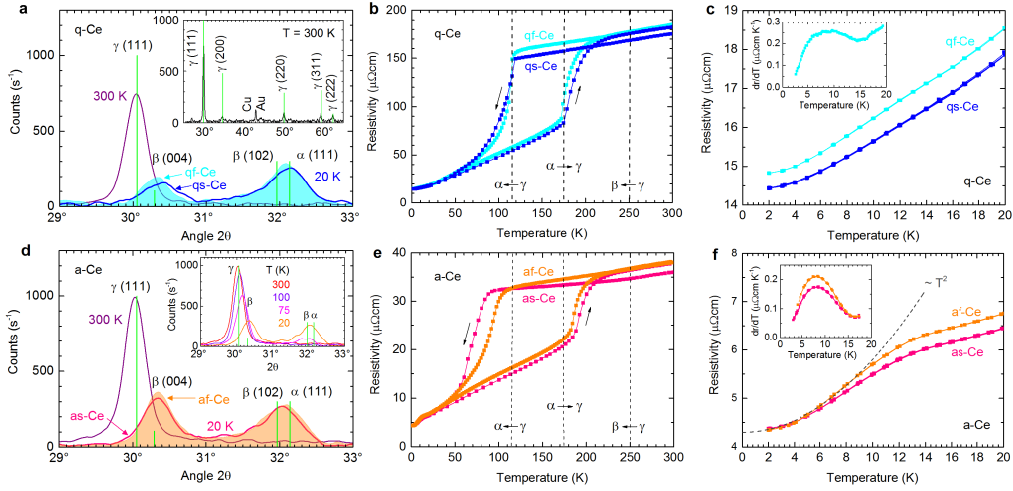}
\caption[]{\textbf{Characterization of Ce samples with different fractions of $\alpha$-Ce. a,} X-ray diffractogram of q-Ce (quenched from the melt) at 300 K and 20 K. Coloured area indicates the intensity obtained after fast cooling from 300 K to 20 K (qf-Ce), solid lines indicate intensities measured after slow cooling (qs-Ce). Green bars mark reflections according to the ``International Centre for Diffraction Data'' (ICDD) data base for $\alpha$-Ce (No. 78-0640), $\beta$-Ce (No. 89-2728), and $\gamma$-Ce (No. 78-0638). The inset shows the full diffractogram at 300 K, where ``Cu'' and ``Au'' indicate contributions arising from the sample holder. \textbf{b,} Resistivity cycles vs. temperature of q-Ce for fast and slow cooling and heating. Dashed vertical lines indicate the transition temperature for the $\gamma \rightarrow \beta$, $\gamma \rightarrow \alpha$, and $\alpha \rightarrow \gamma$ phase transitions according to Ref. \cite{gschneidner_effects_1962}. \textbf{c,} Resistivity of the second cooling cycle. The inset shows the derivative $d \rho/dT$ vs. temperature. \textbf{d-f,} X-ray diffraction and resistivity data of sample a-Ce (annealed) similar to \textbf{a-c}. The inset of \textbf{d} shows the reflections for different temperatures. The dashed line in \textbf{f} indicates a $\rho \propto T^2$ behaviour. The inset shows $d \rho/dT$ vs. $T$.}
\label{fig2}
\end{figure}

In order to investigate the effect of the 4$f$ configuration on single-atom Ce contacts, we employed the content of $\alpha$-Ce in our Ce samples as a control parameter. To this end, we subjected the Ce ingots to different heat treatments, see Methods section for details. q-Ce samples had been quenched from the melt to room temperature. These samples had passed very quickly through the $\gamma-\beta$ transition and are expected to contain a large volume fraction of $\alpha$-Ce at low $T$. a-Ce samples had been annealed at 600 $^{\circ}$C and slowly cooled to 100 $^{\circ}$C. The $\gamma-\beta$ transition ($T_{\gamma \beta}$ = 60 $^{\circ}$C) was passed even more slowly to 40 $^{\circ}$C at a rate of 3 $^{\circ}$C/h. These samples are expected to contain a larger volume fraction of $\beta$-Ce and a smaller fraction of $\alpha$-Ce at low temperatures. The samples were then cooled from room temperature to low temperatures and their structure was checked by x-ray diffraction shown in Fig. \ref{fig2}.

At room temperature $T$ = 300 K all Bragg reflections can be assigned to the $\gamma$ phase of cerium, see inset of Fig. \ref{fig2}a. The intensities of the individual Bragg reflections deviate from the intensities expected for a powder of randomly distributed grains due to a preferred orientation of some crystallites along the [111] direction in the polycrystalline ingot. Upon cooling to 20 K, the $\gamma$(111) reflection shifts to larger angles (smaller lattice-plane distances) and is at 20 K attributed to the $\beta$(004) reflection. Below 100 K an additional peak develops around $2 \theta = 32^{\circ}$, see inset of Fig. \ref{fig2}d, due to the transformation to the $\alpha$ phase below 100 K. Figs. \ref{fig2} a and d clearly show that a phase mixture of $\alpha$-Ce and $\gamma$-Ce forms at low temperatures with a higher fraction of $\alpha$-Ce in q-Ce compared to a-Ce. Furthermore, the fraction depends on whether the sample was cooled fast (qf-Ce, af-Ce, coloured areas) or slowly (qs-Ce, as-Ce, solid lines). For a quantitative estimate of the fraction of $\alpha$-Ce at 20 K, we estimate the ratio $S_{\alpha}$ of the integrated intensities $I$ of the two peaks centered at $2 \theta = 30.3^{\circ}$ and $32.1^{\circ}$, $S_{\alpha} = I(32.1^{\circ})/[I(32.1^{\circ})+I(30.3^{\circ})]$. Table \ref{table1} shows that the fast cooled qf-Ce has a much larger $S_{\alpha}$ and, hence, fraction of $\alpha$-Ce than the slowly cooled as-Ce. These results are in perfect agreement with earlier investigations by Gschneidner et al. \cite{gschneidner_effects_1962}.
  
The phase transformations are also observed in the temperature dependence of the resistivity, see Figs. \ref{fig2}b and e, for which the annealed a-Ce exhibits a factor-of-five lower resistivity than q-Ce which was rapidly cooled from the melt. In both samples, the resistivity drops while cooling through the $\gamma \rightarrow \alpha$ transition and strongly increases while heating through the $\alpha \rightarrow \gamma$ transition with a hysteresis characteristic for a first-order phase transition. For q-Ce, the temperatures where the resistive transitions set in, agree again very well with data by Gschneidner et al. \cite{gschneidner_effects_1962}. In contrast, the hysteresis observed for a-Ce is much broader, possibly due to the larger amount of $\beta$ phase hampering and delaying the transformation from $\gamma$- to $\alpha$-Ce \cite{james_resistivity_1952}. 

At temperatures below 20 K, a-Ce shows a kink in $\rho(T)$ around 12 K (Fig. \ref{fig2}f) characteristic for the onset of antiferromagnetic order in $\beta$-Ce \cite{james_resistivity_1952,wilkinson_neutron_1961}. In the antiferromagnetic regime, the resistivity follows a $\rho \propto T^2$ dependence due to antiferromagnetic spin waves \cite{ueda_electrical_1977}. The kink is only weakly observed in q-Ce (Fig. \ref{fig2}c). We use the resistivity ratio $RR_{\beta} = \rho(15\, {\rm K})/\rho(2\, {\rm K})$ to indicate the amount of $\beta$ phase in the sample. Table \ref{table1} shows that the relative amount of $\beta$ phase decreases from slowly cooled as-Ce to fast cooled qf-Ce in agreement with the corresponding increase of the fraction of $\alpha$-Ce estimated from the x-ray intensity ratio $S_{\alpha}$. In summary, samples with different volume fractions of $\alpha$-Ce and $\beta$-Ce at low temperatures have been successfully prepared and characterized by x-ray diffraction and resistivity data to disentangle their respective contributions to the total conductance of Ce atomic contacts.

\begin{figure}
\includegraphics[width=\columnwidth,clip=]{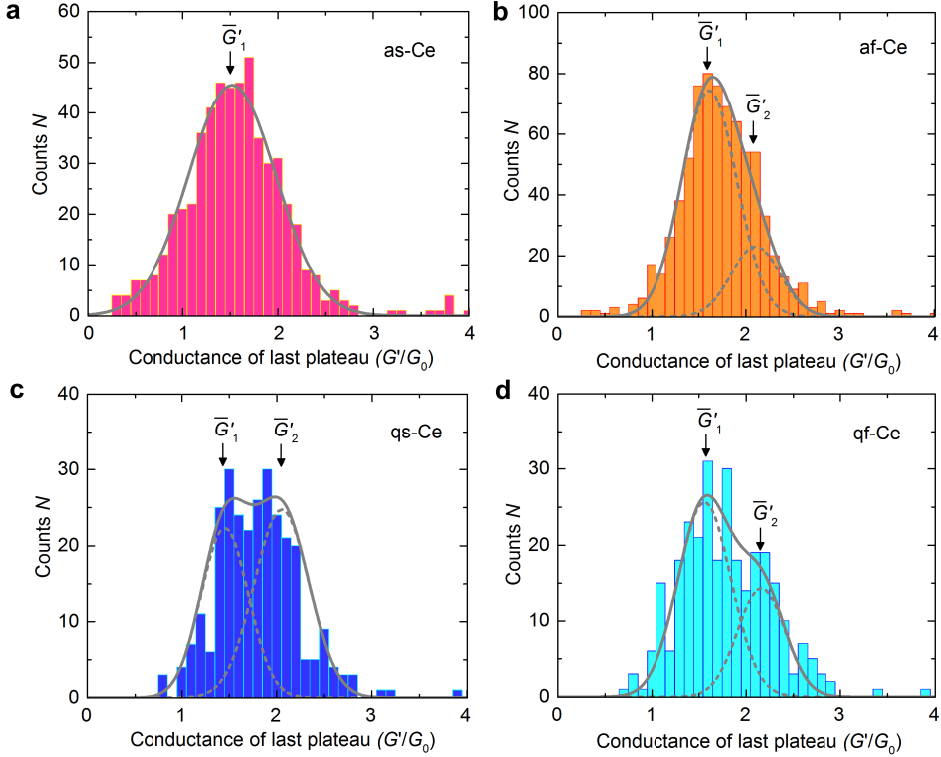}
\caption[]{\textbf{Histograms of the conductance $G'$ of the last plateau of differently prepared Ce MCBJs.} \textbf{a} and \textbf{b} show data obtained for a-Ce samples cooled slowly or fast, respectively, from room temperature to 4.2 K before the measurements were performed. \textbf{c} and \textbf{d} show data obtained in the same manner for q-Ce. For slowly cooled as-Ce the distribution can be described by a single Gauss curve (solid line). In all other cases the data are best described by two Gauss curves (dashed lines), one centered at lower conductance $\bar{G'_1}$, the other at higher conductance $\bar{G'_2}$. The solid lines indicate the sum of the two Gauss curves.}
\label{fig3}
\end{figure}

We now turn to the conductance properties of the different Ce nanocontacts. We proceed by considering the conductance value of the last plateau $G'$ (Fig. \ref{fig1}b) and plot in Fig. \ref{fig3} histograms of four representative samples thus focusing on single-atom-contact histograms instead of those of full conductance curves $G(\Delta x)$. These histograms comprise opening curves only because closing curves generally show much larger last-plateau values extending up to 5\,$G_0$. This suggests that upon closing the contacts instantaneously several atoms form the contact. 

We first note that the conductance values $G'$ of the a-Ce samples follow a Gaussian distribution 
\begin{equation}
N(G') = \frac{A}{\sigma \sqrt{2\pi}}\,{\rm exp}[\frac{-(G'-\bar{G'})^2}{2\sigma ^2}]
\label{eqn1}
\end{equation}
($A$: area, $\bar{G'}$: mean value, $\sigma$: standard deviation) much more closely and smoothly than the q-Ce samples. As mentioned above, the former contain more $\beta$ phase with  localized stable Ce$^{3+}$ moments. It is highly plausible that the local environment of atoms in the contact with reduced number of nearest neighbours leads to a further stabilization of localized moments. This observation suggests that it is not the hybridization of 4$f$ electrons with intra-atomic $6s/5d$ orbitals but rather the hybridization with nearest-neighbour orbitals which is essential. The hybridization possibly further decreases while opening the contact due to the elongation of interatomic bonds when a single atom forms the contact. Indeed, scanning tunneling spectroscopy on single Ce or Co impurities on the Ag or Cu surface show a strong reduction of the Kondo temperature $T_{\rm K}$ for thin layers and for single atoms and clusters as compared to $T_{\rm K}$ of the corresponding bulk solids, due to the reduction of the number of nearest neighbours and the ensuing decrease of hybridization of the magnetic impurity with respect to the bulk electronic system of the host crystal \cite{li_kondo_1998,schneider_kondo_2005,ternes_spectroscopic_2009}. Vice versa, for individual Co adatoms on Cu(100), $T_{\rm K}$ strongly increases toward the bulk value upon decreasing the tip-adatom distance due to the stronger hybridization between the Co 3\textit{d} level and the conduction-electron states of the Cu substrate and the W tip \cite{neel_conductance_2007}. In line with these arguments a substantial change of the 4$f$ electronic structure at the surface of  $\alpha$-Ce towards a $\gamma$-like behaviour was observed in photoemission experiments \cite{weschke_surface_1991}. Therefore, in the as-samples with a large fraction of $\beta$-Ce and stabilization of Ce$^{3+}$ moments in the contact region, a homogeneous distribution of $G'$ is expected. 

The af sample is actually better described by a dominant Gaussian centered at $\bar{G'_1}$ = 1.6 $G_0$ and a three-times smaller Gaussian centered at $\bar{G'_2}$ = 2.1 $G_0$ indicating that fast cooling from room temperature also favours the $\alpha$-phase formation to some extent. On the other hand, the histograms of the q samples can be clearly decomposed into two Gaussians, one centered around $\bar{G'_1} \approx$\, 1.5 $G_0$ and the other around $\bar{G'_2}$ = 2.1 $G_0$. These two components would then correspond to stable Ce moments reminiscent of the $\beta$ phase ($G'_1$) and strongly hybridized moments as in the $\alpha$ phase ($G'_2$) with a finite transmission probability due to their more delocalized nature, respectively. Table \ref{table1} summarizes this behaviour, i.e., samples with a larger volume fraction of $\alpha$-Ce, represented by a large $S_{\alpha}$ and a small $RR_{\beta}$, more frequently show a higher conductance $G'_2$ of the last plateau inferred from the ratio of the areas $A_1$ and $A_2$ of the two distribution functions. The enhanced conductance clearly depends on the sample treatment and the different volume fraction of electronically different phases.
% rather than on shifting electronic levels by built-in electric fields of molecular break junction \cite{champagne_mechanically_2005}.  

We now discuss the implication of our results for the $\gamma - \alpha$ transition. In both KVC and MT models the \textit{f} electrons are strongly correlated in the $\alpha$ and $\gamma$ phases and both models are in qualitative agreement with the localization-delocalization picture. Specifically, the MT model of \textit{f} electrons considers localized nonbonding \textit{f} states in $\gamma$-Ce which are favoured by an on-site \textit{f-f} Coulomb interaction $U$ being larger than the \textit{f}-hybridization energy \cite{johansson1974}. While $U$, being a intratomic quantity, might be considered the same for bulk Ce and single-atom Ce contacts, the latter may be strongly reduced via the decrease of 4\textit{f}-(\textit{sd})$^3$ hybridization, see below. Furthermore, the effectively one-dimensional contact would entail an additional reduction of the bandwidth which, in a tight-binding model, is proportional to the number of nearest neighbours \cite{ashcroft_mermin}, i.e., two in the single-atom contact vs. twelve in bulk fcc cerium. This would lead to a strong tendency of pushing the contact towards the Mott-insulating side compared to bulk. Thus, our observation of nearly delocalized 4\textit{f} electrons participating in electronic conduction is at variance with the MT model for the $\gamma - \alpha$ transition in the contact region.
%When the volume decreases, the \textit{f-f} and \textit{f-sd} hopping increases and the energy gain by building Bloch states becomes larger than the increase in Coulomb energy. Hence, itinerant \textit{f} states participate in the bonding in $\alpha$-Ce. 

In the Kondo volume collapse (KVC) model, the 4\textit{f} electrons are nearly localized and exhibit a stable moment in both $\alpha$- and $\gamma$-Ce but experience a different screening by the \textit{sd} conduction electrons resulting in unscreened moments in $\gamma$-Ce and screened moments in $\alpha$-Ce. Spin fluctuations give rise to the phase transition \cite{lavagna_volume_1982,allen_kondo_1982}. The KVC is corroborated by x-ray diffraction and x-ray emission spectroscopy \cite{lipp_thermal_2008,lipp_x-ray_2012} and has been predicted to occur at the nanoscale down to the dimer level \cite{casadei_density-functional_2012}. Experiments and first-principles calculations suggest that Ce has a low-temperature critical point at negative pressures \cite{thompson_two_1983,lashley_tricritical_2006} and is therefore close to being quantum critical \cite{lanata_2013}. The latter work has pointed out that spin-orbit coupling plays an important role in hampering the local fluctuations induced in the $f$ local space of the large-volume $\gamma$ phase.  

One would expect that, since the 4$f$-($sd$)$^3$ hybridization is primarily with neighbouring atoms, the Kondo temperature $T_{\rm  K} \approx 790$\,K for $\alpha$-Ce \cite{allen_kondo_1982} would be strongly reduced. However, Kondo-like behaviour might still be observed as long as $T_{\rm K}$ exceeds the measuring temperature of 4.2 K. It is important to point out that in one-dimensional metals the Kondo effect leads to an \textit{enhancement} of the conductance as shown in numerous examples \cite{iqbal_odd_2013,park_coulomb_2002,liang_kondo_2002,wiel_kondo_2000}. We are therefore led to the conclusion that although the hybridization between conduction electrons and 4$f$ electron at the single atom of the contact very likely is reduced, the $\alpha$-phase-rich q-Ce contacts facilitate $f$-electron transport across the junction which is qualitatively in line with the KVC model. \\

\vspace{6mm}
\noindent
\textbf{Methods}\\
All samples were prepared from the same Ce starting material (purity 99.99 \%, Atlantic Equipment Engineers). This was first melted several times in an arc furnace under argon atmosphere to obtain a homogeneous ingot. After heating, the melt solidified rapidly by cooling to 18 $^\circ$C in approximately one minute by contact with a water cooled copper plate. The material prepared in this manner was labeled q-Ce. The ingot was cut into two halves one of which was thermally annealed. For this purpose, it was put in an alumina crucible and sealed in a quarz tube under argon atmosphere ($p\approx 5\cdot 10^{-2}$ mbar). The quartz tube was  thermally annealed for several days. First, the temperature was raised to $T=600\,^\circ$C to cross the $\gamma$-$\beta$ phase boundary and then held there for 8 hours. Afterwards the temperature was lowered over a period of about 4 days to $T=100\,^\circ$C and then lowered even more slowly, with a cooling rate of $\partial T / \partial t=3\,^\circ\textnormal{C}/\textnormal{h}$, while crossing the $\gamma$-$\beta$ phase boundary ($T_{\gamma,\beta}\approx 60\,^\circ$C). The material prepared in this manner was labeled a-Ce.\\

X-ray diffraction was done using a Siemens D500 powder-diffractometer equipped with a $^4$He flow cryostat and with the sample under high vacuum ($p=10^{-7}$ mbar). Pieces of 0.5-mm thickness were cut from the ingot in several arbitrary spatial directions. They were properly cleaned in acetone and their oxide layers were carefully removed with a scalpel and then polished abrasive sand paper. Afterwards they were immediately covered with a thin film of highly diluted GE varnish to protect them against further oxidation. X-ray diffractograms were obtained in $\theta$-$2\theta$ Bragg-Brentano mode using Cu-K$_\alpha$ radiation and a Ni foil to reduce the contribution from Cu-K$_\beta$ radiation.\\

Resistivity data were taken on bulk Ce samples, having been subjected to the different heat treatments described above, in a physical-property measurement system (PPMS, Quantum Design) in a four-point probe setup. Cerium pieces of 0.5 mm $\times$ 0.5 mm $\times$ 10 mm size were cut off the ingots and contacted with copper wires using a conductive epoxy EPO-TEK H20E.\\

For the MCBJ, thin wires of $0.1 \times 0.1$\, mm $^2$ cross section and 8 mm length were cut from the ingot. A notch was cut in the middle of the wire as a predetermined point where it should break during bending. The wire was glued with Stycast epoxy to a flexible 0.3-mm thick copper-bronze substrate coated with a 2-$\mu$m thick durimide film
for electrical insulation. It was crucial to heavily coat the sample with Stycast for stabilization of the wire sustaining the structural $\gamma \rightarrow \alpha$ phase transition (with intervening $\beta$ phase) which is accompanied by a huge volume change and thus generation of mechanical stress. The Ce wire was connected with conductive epoxy EPO-TEK H20E to four copper leads in order to perform four-point conductance measurements. This assembly was then mounted in a MCBJ device with countersupports 8 mm apart and cooled to 4.2 K in a $^4$He bath cryostat. The substrate was bent mechanically by pushing a piston against the back of the substrate and fine tuning of the bending was achieved by using a piezo stack controlled by a voltage $V_{\rm p}$. A voltage of 10 $\mu$V was applied to the junction and the current through the junction was measured. The electrode distance $\Delta x$ was obtained from $G(V_{\rm p})$ in the tunneling regime by using appropriate work functions of the materials as described earlier \cite{muller_switching_2011}. All conductance measurements on the MCBJ devices were carried out at 4.2 K.\\

\vspace{6mm}
\noindent
\textbf{Acknowledgements}\\
We thank K. Held and J. Schmalian for valuable discussions and W. Kittler for help with the annealing of the Ce ingot.\\

\clearpage
\begin{table}
\begin{center}
\begin{tabular}{|c|c|c|c|c|c|c|}
  \hline 
  &&&&&&\\
  	Sample &Cooling rate &$S_{\alpha}$&$RR_{\beta}$&$\bar{G'_1}\pm \sigma_1$&$\bar{G'_2}\pm \sigma_2$& $A_2/A_1$  \\
	  &(K min$^{-1})$&  & &($G_0$)&($G_0$)& \\
	\hline
	as-Ce & 1 & 0.49 & 1.62&$1.52\pm 0.46$& - & -\\
	af-Ce & 10 & 0.51 & 1.59&$1.61\pm 0.29$&$2.10 \pm 0.26$& 0.28\\
	qs-Ce & 1 & 0.65 & 1.47&$1.45 \pm 0.25$&$2.05 \pm 0.29$& 1.29\\
	qf-Ce & 10 & 0.67 & 1.39&$1.56 \pm 0.28$&$2.16 \pm 0.25$& 0.50\\
	\hline 
	\end{tabular}
	\caption{\label{table1}\textbf{Characteristic parameters of single-atom Ce contacts.} Ratio $S_{\alpha}$ of integrated x-ray intensities and resistivity ratio $RR_{\beta}$ for the different Ce samples cooled with different rates from 300 K to 20 K. $\bar{G'_1}$ and $\bar{G'_2}$ are the mean values of the two Gaussian distribution functions and $A_2/A_1$ is the ratio of their areas.}
\end{center}
\end{table}

\clearpage

%\clearpage

%\clearpage

\end{document}